\documentclass[twocolumn,twoside,slac_two]{revtex4}
\usepackage{amsmath}
\usepackage{graphicx}
\usepackage{fancyhdr}
\def\OMIT#1{}

\newcommand{\nn}{\nonumber} 

\newcommand{\bn}{{\bar n}}
\newcommand{\bea}{\begin{eqnarray}}
\newcommand{\eea}{\end{eqnarray}}

\newcommand{\SCETa}{\mbox{${\rm SCET}_{\rm I}$ }}

\def\lqcd{\Lambda_{\rm QCD}}

\begin{document}

\title{Hadronic $B$ decays from SCET}

%

\author{Christian W.~Bauer}
\affiliation{Ernest Orlando Lawrence Berkeley National Laboratory and
University of California, Berkeley, CA 94720}

\begin{abstract}
In this talk I will discuss non-leptonic $B$ decays, in particular how soft-collinear effective field theory (SCET) can be used to constrain the non-perturbative hadronic 
parameters required to describe the various observables. 
\end{abstract}

\maketitle

\thispagestyle{fancy}


\section{Introduction}
The standard model (SM) of particle physics has proven to hold up against any experimental
tests it has been subjected to so far. The SM has several striking features for which
no underlying principle has been experimentally confirmed to this date. First, the SM requires
the scale of electro-weak symmetry breaking to be of order a few hundred GeV, which is many orders 
of magnitude below the only fundamental scale of nature we know of, the Planck scale. Second, 
to explain the masses and flavor violating transitions of fermions requires the fundamental Yukawa
matrices to satisfy a very particular scaling, for which no satisfactory symmetry or other underlying principle has been found so far. While the scale of electro-weak symmetry breaking is known from 
the measured properties of gauge interactions, the scale of flavor violation could be completely 
unrelated to that scale. However, many models of new physics which address the 
electro-weak scale also give additional contributions to flavor physics. Thus, precise measurements
of flavor and CP violating observables can severely constrain possible models of 
electro-weak symmetry breaking. 

Since the standard model predicts the short distance couplings of quarks to one another, 
while experimental
measurements are done with hadrons, one needs to understand long distance QCD effects 
on the measured quantities to extract the underlying physics. It is the purpose of this talk 
to discuss how this separation between long and short distance physics can be achieved 
using effective theories. The effective theory that is applicable to the non-leptonic $B$ decays
to two light mesons, as we are concerned with here, is the soft-collinear effective theory 
(SCET)~\cite{SCET}.

\section{Factorization in $B \to M_1 M_2$}
There has been tremendous progress over the last few years in understanding
charmless two-body, non-leptonic $B$ decays in the heavy quark limit of
QCD~\cite{QCDF,PQCD,pipiChay,bprs,Bauer:2001cu,pQCDKpi,BW}.
In this limit one can prove factorization theorems of the matrix elements
describing the strong dynamics of the decay into simpler structures such as
light cone distribution amplitudes of the mesons and matrix elements describing
a heavy to light transition~\cite{QCDF}. It is very important that these results are obtained from
a systematic expansion in powers of $\lqcd/m_b$. The development of
soft-collinear effective theory (SCET)~\cite{SCET} allowed these decays to be
treated in the framework of effective theories, clarifying the separation of
scales in the problem, and allowing factorization to be generalized to all
orders in $\alpha_s$. 

Factorization for $B\to M_1 M_2$ decays involves three distinct distance scales
$m_b^2 \gg E_M\Lambda \gg \Lambda^2$. For $B\to M_1 M_2$ decays, a factorization
theorem was proposed by Beneke, Buchalla, Neubert and Sachrajda~\cite{QCDF},
often referred to as the QCDF result in the literature. Another proposal is a
factorization formula which depends on transverse momenta, which is referred to
as PQCD~\cite{PQCD}. The factorization theorem derived using
SCET~\cite{bprs,pipiChay} agrees with the structure of the QCDF proposal if
perturbation theory is applied at the scales $m_b^2$ and $m_b\Lambda$.  One of the differences is that QCDF
treats $c\bar c$ penguins perturbatively, while in the SCET analysis they are
left as a perturbative contribution plus an unfactorized large ${\cal O}(v)$
term.  The SCET result improved the factorization formula by
generalizing it to allow each of the scales $m_b^2$, $E_M\Lambda$, and
$\Lambda^2_{\rm QCD}$ to be discussed independently.  

The derivation of the SCET factorization theorem occurs in several steps, corresponding 
to integrating out the various scales in the problem. As already mentioned, the  relevant scales are $\mu \sim m_b$, 
$\mu \sim \sqrt{m_b \lqcd}$ and $\mu \sim \lqcd$. One starts from the effective
weak Hamiltonian, which describes the effects of the weak physics in terms of local 4-quark operators. 
Integrating out $\sim m_b$ fluctuations, the effective Hamiltonian at leading order in
\SCETa\cite{bps4} can be written schematically as
\begin{eqnarray} \label{match}
 H_W \!\!&=  &\!\! \frac{2G_F}{\sqrt{2}}  \sum_i \left[
       c_i\otimes  Q_{i}^{(0)}
  + b_i\otimes
  Q_{i}^{(1)}
  + {\cal Q}_{c\bar c} \right] \,,
\end{eqnarray}
where $c_i$ and $b_i$ are Wilson coefficients, and the symbol $\otimes$ denotes 
convolutions over various momentum fractions. 

\begin{figure}[t]
\centering
\includegraphics[width=70mm]{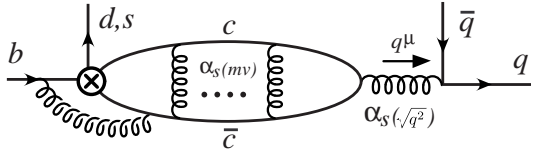}
\caption{Example of long distance charming penguins. The $mv$ gluons
      are nonperturbative and LO soft gluons are exchanged by the $b$, $c$, $\bar
      c$ and spectator quark which is not shown. } \label{fig:cpenguin}
\end{figure}
The term ${\cal Q}_{c\bar c}$
denotes operators appearing in long distance charm effects as in
Fig.~\ref{fig:cpenguin}. There is broad agreement that
charm loop contributions from hard ($\sim\! m_b$) momenta can be computed in
perturbation theory and they are included in the $c_i$ and $b_i$ coefficients in Eq.~(\ref{match}). However, there are non-perturbative
contributions from penguin charm quark loops (so-called charming
penguins~\cite{cpens}), which are contained in ${\cal Q}_{c\bar c}$. 
While no proof of factorization for the matrix element of this operator exists, it
is still possible to determine its parametric dependence on
$m_c/m_b$, $v$, and $\Lambda_{\rm QCD}/m_b$ using operators in effective field
theories. We find
\begin{eqnarray} \label{Acc}
 \frac{A_{c\bar c}^{\pi\pi}}{A^{\pi\pi}_{LO}} \sim
  \alpha_s(2m_c) \: f\Big(\frac{2m_c}{m_b}\Big)\: v 
 \,,
\end{eqnarray}
Thus, this contribution gives rise to a source of strong phases in the amplitudes, while
all strong phases vanish for the other terms. For this reason, we will keep this term and treat 
its matrix element as an unknown complex parameter in the theory. 

The collinear fields in the operators $O_i^{(0,1)}$ can be decoupled from the ultrasoft fields by 
making a simple field redefinition~\cite{SCET}. The operators $O_i^{(0,1)}$ then factor into $(n,v)$ and $\bn$
parts,
\begin{eqnarray} \label{split}
 Q_{i}^{(0,1)} = \tilde Q_{i}^{(0,1)} Q_{i}^\bn \,.
\end{eqnarray}
The first term contains both a soft $b$ quark field and a collinear field in the $\bn$ direction, while the second term contains two collinear fields in the $n$ direction. The matrix element of the operators $O_i^{(0,1)}$ thus factor into a $B \to M$ transition matrix element, and a vacuum to light meson matrix element. 

Putting all these results together, we obtain the SCET factorization formula 
\begin{eqnarray}  \label{A0newfact}
A \!\!&=&\!\! 
\frac{G_F m_B^2}{\sqrt2}\! \bigg[ \bigg\{
   f_{M_1}\! \int_0^1\!\!\!\!du\, dz\,
    T_{1\!J}(u,z) \zeta^{BM_2}_{J}(z) \phi^{M_1}(u) 
   \nn \\
 &&\hspace{0.0cm}
   + f_{M_1} \zeta^{BM_2}\!\! \int_0^1\!\!\!\! du\, T_{1\zeta}(u) \phi^{M_1}(u)
  \bigg\} \!+\! \Big\{ 1\leftrightarrow 2\Big\} \nn\\
 && \hspace{0.0cm} + \lambda_c^{(f)} A_{c\bar c}^{M_1M_2} \bigg] ,  
\end{eqnarray}
where $\zeta^{BM}$ and $\zeta^{BM}_J$ are non-perturbative parameters describing
$B \to M$ transition matrix elements, and $A_{\rm c\bar c}^{M_1 M_2}$
parameterizes complex amplitudes from charm quark contractions for which
factorization has not been proven. Power counting implies $\zeta^{BM}\sim
\zeta^{BM}_J\sim (\Lambda/Q)^{3/2}$.  $T_{1\!J}(u,z)$ and $T_{1\zeta}(u)$ are
perturbatively calculable in an expansion in $\alpha_s(m_b)$ and depend upon the
process of interest.

At leading order in $\alpha_s(m_b)$ the short distance coefficients $T_{1\!J}(u,z)$ is independent
of the parameter $z$, which implies that the functional form of the non-perturbative function $\zeta_J^{BM}(z)$ does not matter, since we can define a new hadronic parameter $\zeta_J^{BM} \equiv \int \!dz\, \zeta_J^{BM}(z)$. 

\section{Phenomenology}

\subsection{Counting of hadronic parameters}
Without any theoretical input, there are 4 real hadronic parameters for each
decay mode (one complex amplitude for each CKM structure) minus one overall
strong phase.  In addition, there are the weak CP violating phases that we want
to determine. For $B \to \pi \pi$ decays there are a total of 11 hadronic
parameters, while in $B \to K \pi$ decays there are 15 hadronic parameters.

Using isospin, the number of parameters is reduced. Isospin gives one amplitude
relation for both the $\pi \pi$ and the $K \pi$ system, thus eliminating 4
hadronic parameters in each system (two complex amplitudes for each CKM
structure). This leaves 7 hadronic parameters for $B \to \pi \pi$ and 11 for $B
\to K \pi$. 

The SU(3) flavor symmetry relates not only the decays $B \to \pi \pi$ and $B \to
K \pi$, $B  \to K K$, but  also $B \to \pi \eta_8$,  $B \to \eta_8 K$  and $B_s$
decays to  two  light mesons. The  decomposition  of the amplitudes  in terms of
SU(3)     reduced          matrix        elements     can       be      obtained
from~\cite{Zeppenfeld,SavageWise,GrinsteinLebed}. and  20 hadronic
parameters are required  to describe all  these decays minus 1  overall phase (plus additional
parameters for singlets and mixing to properly describe $\eta$ and $\eta'$).  Of
these hadronic parameters, only 15 are required to describe $B  \to \pi \pi$ and
$B \to K \pi$ decays (16 minus an overall phase).  If we add $B \to K K $ decays
then 4  more   paramaters  are needed  (which  are  solely  due to   electroweak
penguins).  

\begin{table}
\begin{tabular}{l||c|c|c|c|c|}
& no & &  & SCET & SCET   \\[-2pt]
 & expn. &  \raisebox{1.6ex}[0pt]{SU(2)} 
 & \raisebox{1.6ex}[0pt]{SU(3)} 
 & +SU(2)&+SU(3) \\\hline
$B \to \pi \pi$  & 11 & 7/5 & & 4 &\\ \cline{1-3}\cline{5-5}
$B \to K \pi$  & 15& 11 & \raisebox{1.6ex}[0pt]{15/13} &
    +5(6) & \raisebox{1.6ex}[0pt]{4} \\ \hline
$B \to K \bar K$  & 11 & 11 & +4/0 & +3(4) & +0
\end{tabular}
\caption{
Number of real hadronic parameters from different expansions in QCD. The first column
shows the number of theory inputs with no approximations, while the next columns
show the number of parameters using only SU(2), 
using only SU(3), using SU(2) and SCET, and using SU(3) with SCET.  For the 
cases with two numbers, $\#/\#$, the second follows from the first after 
neglecting the small penguin coefficients, ie setting $C_{7,8}=0$.  In SU(2) 
+ SCET $B\to K\pi$ has 6 parameters, but 1 appears already in $B\to \pi\pi$, 
hence the $+5(6)$. The notation is analogous for the $+3(4)$ for 
$B\to K\bar K$. 
\label{table_parameters}
}
\end{table}
The number of parameters that occur at leading order in different expansions of
QCD are summarized in Table~\ref{table_parameters}, including the SCET
expansion.  The parameters with isospin+SCET are
\begin{eqnarray} \label{params}
\pi\pi: & & \{\zeta^{B\pi}\!+\! \zeta_J^{B\pi},\beta_\pi  \zeta_J^{B\pi},
    P_{\pi\pi} \} \,, \\
K\pi:   && \{\zeta^{B\pi}\!+\! \zeta_J^{B\pi},\beta_{\bar K}  \zeta_J^{B\pi},
    \zeta^{B\bar K}+\zeta_J^{B\bar K}, \beta_\pi \zeta_J^{B\bar K}, 
     P_{K\pi} \} \,, \nn \\
K\bar K: && \{\zeta^{B\bar K}+ \zeta_J^{B\bar K},\beta_K \zeta_J^{B\bar K}, 
     P_{K\bar K} \} \,. \nn
\end{eqnarray}
Here $P_{M_1 M_2}$ are complex penguin amplitudes and the remaining parameters
are real.

Taking SCET + SU(3) we have the additional relations $\zeta^{B\pi} = \zeta^{B K}
= \zeta^{B\bar K}$, $\zeta_J^{B\pi} = \zeta_J^{B K} = \zeta_J^{B\bar K}$,
$\beta_\pi=\beta_K=\beta_{\bar K}$, and $A_{cc}^{\pi\pi}= A_{cc}^{K\pi} =
A_{cc}^{K\bar K}$ which reduces the number of parameters considerably.

\subsection{Implications of small phases}
In SCET, the only source of strong phases are from the charm penguin amplitude parameter $A_{cc}$. 
Since by SU(2) flavor symmetry there is only a single such amplitude parameter for the decays $B \to K \pi$, all relative strong phases between $A_{cc}$ and any other term are the same, while relative 
strong phases between any other two amplitude parameters are identically zero. This result can be used to make several predictions in SCET, which are relatively independent of the actual size of the value of the amplitude parameters. An example are certain sum rules in the decays $B \to K \pi$, which are constructed out of the ratios of brancing ratios
\begin{eqnarray}
  R_1 &=& \frac{2 {\rm Br}(B^-\to \pi^0K^-)}
     {{\rm Br}(B^- \to \pi^- \bar K^0)} -1 
     \,,\\
    R_2 &=& \frac{ {\rm Br}(\bar B^0\to \pi^-K^+)\tau_{B^-}}
     {{\rm Br}(B^- \to \pi^- \bar K^0)\tau_{B^0}} -1
     \,,\nn\\
    R_3 &=& \frac{2 {\rm Br}(\bar B^0\to \pi^0\bar K^0)\tau_{B^-}}
     {{\rm Br}(B^- \to \pi^- \bar K^0)\tau_{B^0}} -1
    \,,\nn
\end{eqnarray}
and rescaled asymmetries
\begin{eqnarray}
  \Delta_1 &=& (1+R_1) A_{\rm CP}(\pi^0 K^-)
     \,,\\
  \Delta_2 &=& (1+R_2)A_{\rm CP}(\pi^- K^+)
     \,,\nn\\
  \Delta_3 &=& (1+ R_3) A_{\rm CP}(\pi^0 \bar K^0)  
    \,,\nn \\
  \Delta_4 &=& A_{\rm CP}(\pi^- \bar K^0) \,. \nn
\end{eqnarray}
They can be combined into linear combinations, which satisfy
\begin{eqnarray}
R_1-R_2+R_3 \sim {\cal O}(\epsilon^2)
\end{eqnarray}
and
\begin{eqnarray}
\Delta_1 - \Delta_2 + \Delta_3 - \Delta_4 \sim \epsilon^2 \sin\gamma \sin(\Delta\phi)
\end{eqnarray}
Here $\epsilon$ denotes a small parameter that is either proportional to $\lambda_u/\lambda_c$ or $C_{9,10}/C_4$. Thus, these combinations of parameters are expected to give contributions which are much smaller than each of the individual terms. Furthermore, the fact that the sum rule for the CP asymmetries is proportional to the differences of strong phases, reduces the predicted result in SCET even more.  
\begin{table*}[t!]
\begin{center}
\caption{Predicted CP averaged branching ratios ($\times 10^{-6}$, first row) and direct CP asymmetries (second row for each mode) for $\Delta S=0$ and $\Delta S=1$ $B$ decays (separated by 
horizontal line) to isosinglet pseudoscalar mesons. The Theory I and Theory II columns give predictions corresponding to two possible solutions for the hadronic parameters. The errors on the predictions are estimates of SU(3) breaking, $1/m_b$ corrections and errors
due to SCET parameters, respectively. No prediction on CP asymmetries is given, if $[-1,1]$ range is allowed at $1\sigma$.}\label{tab1}
\begin{tabular}{llll}
Mode  & Exp.  & Theory I & Theory II\\ \hline
$B^-\to \pi^-\eta$		& $4.3\pm0.5 \;(S=1.3)$ 		& $4.9\pm 1.7\pm 1.0\pm 0.5$		& $5.0\pm 1.7\pm 1.2\pm 0.4$\\
						& $-0.11\pm0.08$	& $0.05\pm 0.19\pm 0.21\pm 0.05$	& $0.37\pm 0.19\pm 0.21\pm 0.05$\\
$B^-\to \pi^-\eta'$		& $2.53\pm0.79 \;(S=1.5)$ 	& $2.4\pm 1.2\pm 0.2\pm 0.4$		& $2.8\pm 1.2\pm 0.3\pm 0.3$\\
						& $0.14\pm0.15$		& $0.21\pm 0.12\pm 0.10\pm 0.14$	& $0.02\pm 0.10\pm 0.04\pm 0.15$\\
$\bar B^0\to \pi^0\eta$	& $<2.5$ 				& $0.88\pm 0.54\pm 0.06\pm 0.42$	& $0.68\pm 0.46\pm 0.03\pm 0.41$\\
						& $-$				& $0.03\pm 0.10\pm 0.12\pm 0.05$	& $-0.07\pm 0.16\pm 0.04\pm 0.90$\\
$\bar B^0\to \pi^0\eta'$& $<3.7$ 				& $2.3\pm 0.8\pm 0.3\pm 2.7$ & $1.3\pm 0.5\pm 0.1\pm 0.3$\\
						& $-$				& $-0.24\pm 0.10\pm 0.19\pm 0.24$ & $-$\\	
$\bar B^0\to \eta\eta$	& $<2.0$ 				& $0.69\pm 0.38\pm 0.13\pm 0.58$ & $1.0\pm 0.4\pm 0.3\pm 1.4$\\
						& $-$				& $-0.09\pm 0.24\pm 0.21\pm 0.04$ & $0.48\pm 0.22\pm 0.20\pm 0.13$\\	
$\bar B^0\to \eta\eta'$	& $<4.6$ 				& $1.0\pm 0.5\pm 0.1\pm 1.5$ & $2.2\pm 0.7\pm 0.6\pm 5.4$\\
						& $-$				& $-$ & $0.70\pm 0.13\pm 0.20\pm 0.04$\\
$\bar B^0\to \eta'\eta'$& $<10$ 				& $0.57\pm 0.23\pm 0.03\pm 0.69$ & $1.2\pm 0.4\pm 0.3\pm 3.7$\\
						& $-$				& $-$ & $0.60\pm 0.11\pm 0.22\pm 0.29$\\\hline
$\bar B^0\to \bar K^0\eta'$	& $63.2\pm4.9 \;(S=1.5)$ 	& $63.2\pm 24.7\pm 4.2\pm 8.1$ & $62.2\pm 23.7\pm 5.5\pm 7.2$\\
							& $0.07\pm0.10 \;(S=1.5)$	& $0.011\pm 0.006\pm 0.012\pm 0.002$ & $-0.027\pm 0.007\pm 0.008\pm 0.005$\\
$\bar B^0\to \bar K^0\eta$	& $<1.9$ 		& $2.4\pm 4.4\pm 0.2\pm 0.3$ & $2.3\pm 4.4\pm 0.2\pm 0.5$\\
							& $-$				& $0.21\pm 0.20\pm 0.04\pm 0.03$ & $-0.18\pm 0.22\pm 0.06\pm 0.04$\\
$B^-\to K^-\eta'$	& $69.4\pm2.7$ 				& $69.5\pm 27.0\pm 4.3\pm 7.7$ & $69.3\pm 26.0\pm 7.1\pm 6.3$\\
					& $0.031\pm0.021$	& $-0.010\pm0.006\pm0.007\pm0.005$& $0.007\pm0.005\pm0.002\pm0.009$\\
$B^-\to K^-\eta$	& $2.5\pm0.3$ 				& $2.7\pm 4.8\pm 0.4\pm 0.3$ & $2.3\pm 4.5\pm 0.4\pm 0.3$\\
					& $-0.33\pm0.17\;(S=1.4)$	& $0.33\pm 0.30\pm 0.07\pm 0.03$ & $-0.33\pm 0.39\pm 0.10\pm 0.04$\\						
\end{tabular}
\label{example_table_2col}
\end{center}
\end{table*}

\begin{figure}[t!]
\begin{center}
\includegraphics[width=8.5cm]{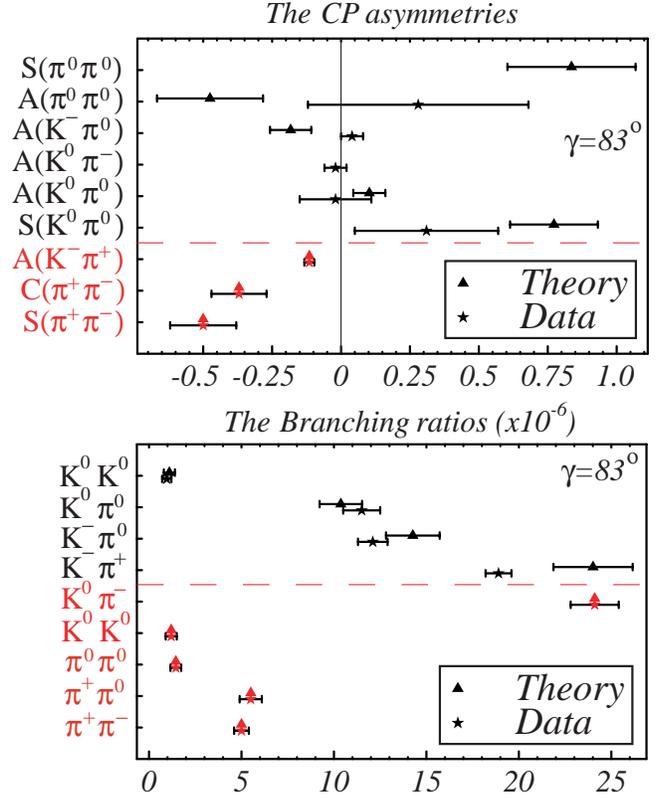}
\caption{Comparison of theory and experiment for all available data in $B \to
  \pi \pi$ and $B \to K \pi$ decays, with $\gamma=83^\circ$. The 8 pieces of data in red
  (below the dashed line) have been used to determine the SCET hadronic parameters 
  $\zeta^{B\pi}$, $\zeta_J^{B\pi}$, $P_{\pi\pi}$     as described in the text. The data above the line are predictions. The CP
  asymmetry in $B^- \to K^0 \pi^-$ is expected to be small, but its numerical 
  value is not predicted reliably.
\label{fig2}}
\end{center}
\end{figure}
Experimentally, one finds
\begin{eqnarray}
R_1-R_2+R_3 = (0.19 \pm 0.15)^{\rm expt}\nn\\
\Delta_1 - \Delta_2 + \Delta_3 - \Delta_4 = ( 0.14 \pm 0.15)^{\rm expt}\,,
\end{eqnarray}
and these results are consistent with zero. The SCET predictions are considerably more precise than 
the current measurements, and using conservative ranges
$\zeta^{B\pi}+\zeta^{B\pi}_J= 0.2\pm0.1$, $\beta_{\bar K}\zeta_J^{B\pi} = 0.10
\pm 0.05$, $\zeta^{B\bar K}+\zeta^{B\bar K}_J= 0.2\pm0.1$, $\beta_\pi
\zeta_J^{B\bar K} = 0.10 \pm 0.05$, $\gamma=70^\circ
\pm 15^\circ$,and all phase differences $\Delta \phi =
0^\circ \pm 30^\circ$ one finds
\begin{eqnarray}\label{sumerrors1}
  R_1 - R_2 + R_3 = 0.028 \pm 0.021 \,,
\end{eqnarray}
and for the CP-sum rule
\begin{eqnarray}\label{sumerrors2}
 \Delta_1 -\Delta_2 + \Delta_3 -\Delta_4 = 0 \pm 0.013 \,.
\end{eqnarray}
Experimental deviations that are larger than these would be a signal for new
physics.

\subsection{General analysis for $B \to \pi \pi$, $B \to K \pi$ decays}
From Table~\ref{table_parameters} and Eq.~(\ref{params}) we can see that a total of 
12 hadronic parameters are required
to describe the decays $B \to \pi \pi$, $B \to K \pi$ and $B \to K K $ at leading order in SCET. On top of that, there is 
one weak phase $\gamma$ which we will take as an unknown parameter. However, in both the decays
$B \to K \pi$ and $B \to KK$, the coefficients multiplying the $B \to K$ transition matrix elements $\zeta^{BK}$ and $\zeta_J^{BK}$ are very small, which implies that the observables are insensitive to the
numerical value of these 
hadronic matrix elements. This eliminates three of the hadronic parameters, namely $\zeta^{BK}+\zeta_J^{BK}$, $\beta_\pi \zeta_J^{BK}$ and $\beta_K \zeta_J^{BK}$. Finally, if we take the
inverse moments of pion and kaon wave functions as experimental input, we obtain one additional relation between hadronic parameters. This leaves us with a total of 7 hadronic parameters as well as 
one weak phase. 
The 8 measurements used 
to fix these parameters are the branching ratios for $B$  decays to $\pi^+ \pi^-$, $\pi^+ \pi^0$, 
$\pi^0 \pi^0$, $K^0 K^0$, $K^0 \pi^-$, as well as the CP asymmetries $S(\pi^+ \pi^-)$, 
$C(\pi^+ \pi^-)$ and $A_{\rm CP}(K^- \pi^+)$. 

Using the hadronic parameters extracted from the $B \to \pi \pi$ decays ($\zeta^{B\pi}$, $\zeta_J^{B\pi}$ and $P_{\pi\pi}$), the
value for $P_{K \pi}$ determined from the decays $B^- \to \pi^- \bar K^0$ and
$\bar B^0 \to \pi^- K^+$ decays and independently varying
$\zeta^{BK}+\zeta^{BK}_J = 0.2\pm 0.1$ and $\beta_\pi \zeta^{BK}_J = 0.10\pm
0.05$, we can calculate all the remaining currently measured $K\pi$ observables.
The results are shown in Fig.~\ref{fig2}. The data used in the fit are shown in red below the dashed
dividing line while those above the line are predictions.  Note that there is
one more piece of data below the line than there are hadronic parameters. This additional 
experimental information was used to determine the value $\gamma = 83^\circ$.

We see that $\gamma=83^\circ$ gives a good match to the $B\to
\pi\pi$ data except for the asymmetry $C(\pi^0\pi^0)$. When taking into account
the theoretical error the most striking disagreements are the ${\rm
  Br}(K^-\pi^+)$ at $2.3\sigma$ and the CP-asymmetry $A_{\rm CP}(K^-\pi^0)$ at
$2.6\sigma$. All other predictions agree within the uncertainties. Note that one could demand that $A_{\rm CP}(K^-\pi^0)$ be reproduced, which would imply a negative value of $\zeta_J^{BK}$ (a naive fit for $\gamma = 83^\circ$ gives $\zeta_J^{BK}\sim -0.15$). Note however, that this would imply that both perturbation theory at the intermediate scale $\mu = \sqrt{E \Lambda}$ and SU(3) are badly broken. 

\subsection{Including isosinglet mesons}

The above analysis has recently been repeated to include decays to iso-singlet final states~\cite{ZW}. This requires adding additional contributions which arise from purely gluonic configurations. It turns out that the additional operators do not change the form of the factorization theorem given in 
Eq.~(\ref{A0newfact}), but the hadronic parameters $\zeta$, $\zeta_J$ and $A_{cc}$ receive
order one contributions from these additional operators. To add isosinglet mesons to the phenomenological analysis thus requires a second set of parameters $\zeta_g$, $\zeta_{J,g}$ and $A_{cc,g}$, which have to be determined from data separately. Since experimentally there are not enough decays available, SU(3) flavor symmetry is required to retain predictive power. At the present time, 
there are two solutions possible for the gluonic hadronic parameters, and the degeneracy can only 
be lifted with further data. The results of the global fit, as taken from~\cite{ZW}, are shown in Table~\ref{tab1}.

\section{Conclusions}
In this talk I have discussed how one can separate the long distance non-perturbative physics from the underlying short distance physics using the soft-colinear effective theory. One finds that the number of 
hadronic parameters is significantly reduced, such that they can be extracted directly from a subset of the data, and then used to make predictions for the remaining data. I have given a brief discussion of the factorization theorem as it emerges from SCET, and then discussed three phenomenological applications. First, I gave a detailed counting of the hadronic parameters using various theoretical approaches, then I discussed a few impacts of the fact that there is only one source of strong phases in the decay amplitudes, and finally, I showed results for global analyses of $B$ decays to two pseudoscalar mesons, with and without including isosinglet mesons.  

\bigskip 

\end{document}